\documentclass[epj]{webofc}\usepackage[varg]{txfonts}\usepackage{bm}
\usepackage{booktabs}\usepackage{color}\woctitle{QCD@Work 2016}
\begin{document}\title{Charmed Mesons and Charmonia:\\Three-Meson
Strong Couplings}\author{Wolfgang Lucha\inst{1}\fnsep
\thanks{\email{Wolfgang.Lucha@oeaw.ac.at}}\and Dmitri Melikhov
\inst{2,3}\fnsep\thanks{\email{dmitri_melikhov@gmx.de}}\and Hagop
Sazdjian\inst{4}\fnsep\thanks{\email{sazdjian@ipno.in2p3.fr}}\and
Silvano Simula\inst{5}\fnsep\thanks{\email{simula@roma3.infn.it}}}
\institute{Institute for High Energy Physics, Austrian Academy of
Sciences, Nikolsdorfergasse 18, A-1050 Vienna, Austria\and
D.~V.~Skobeltsyn Institute of Nuclear Physics, M.~V.~Lomonosov
Moscow State University, 119991, Moscow, Russia\and Faculty of
Physics, University of Vienna, Boltzmanngasse 5, A-1090 Vienna,
Austria\and IPN, CNRS/IN2P3, Universit\`e Paris-Sud 11, F-91406
Orsay, France\and INFN, Sezione di Roma Tre, Via della Vasca
Navale 84, I-00146 Roma, Italy}

\abstract{Revisiting the strong couplings of three mesons, each of
which involves at least one charm quark, proves clear disaccord
between quark-model and QCD sum-rule~results.}\maketitle

\section{Three-meson strong couplings from relativistic
constituent-quark model}\label{Sec:3}We extract the strong
couplings of three mesons among which there is, at least, one of
the charmonia~$\eta_c$ and $J/\psi$ from the residues of poles in
adequate transition form factors for timelike momentum transfer,
which, in turn, are inferred from a relativistic dispersion
approach based on a constituent-quark model.

\section{Definitions of strong couplings, transition form factors,
decay constants}\label{Sec:def}Preparatively, let us introduce all
the quantities necessary for the formulation of the relation
sought,~for \emph{pseudoscalar\/} mesons $P,$ with mass $M_P,$ and
\emph{vector\/} mesons $V,$ with mass $M_V$ and polarization
vector~$\varepsilon_\mu.$\begin{itemize}\item With momentum
transfer $q\equiv p_1-p_2,$ the \emph{strong couplings\/}
$g_{PP'V}$ and $g_{PV'V}$ determine the amplitudes$$\langle
P'(p_2)\,V(q)|P(p_1)\rangle=-\frac{g_{PP'V}}{2}\,(p_1+p_2)^\mu\,
\varepsilon^*_\mu(q)\ ,\qquad\langle V'(p_2)\,V(q)|P(p_1)\rangle=
-g_{PV'V}\,\epsilon_{\varepsilon^*(q)\,\varepsilon^*(p_2)\,p_1\,p_2}\
.$$\item The \emph{transition form factors\/} ${\cal F}(q^2)=
F_+^{P\succ P'}(q^2),$ $V^{P\succ V}(q^2)$ or $A_0^{P\succ
V}(q^2)$ enter in the two-meson matrix elements of the vector
quark current $j_\mu\equiv\bar q_1\,\gamma_\mu\,q_2$ and the
axial-vector quark current $j_\mu^5\equiv\bar
q_1\,\gamma_\mu\,\gamma_5\,q_2$\begin{align}\langle
P'(p_2)|j_\mu|P(p_1)\rangle&=F_+^{P\succ
P'}(q^2)\,(p_1+p_2)_\mu+\cdots\ ,\qquad\langle
V(p_2)|j_\mu|P(p_1)\rangle=\frac{2\,V^{P\succ
V}(q^2)}{M_P+M_V}\,\epsilon_{\mu\,\varepsilon^*(p_2)\,p_1\,p_2}\
,\nonumber\\\langle V(p_2)|j_\mu^5|P(p_1)\rangle&={\rm i}\,q_\mu
\left(\varepsilon^*(p_2)\,p_1\right)\frac{2\,M_V}{q^2}\,A_0^{P\succ
V}(q^2)+\cdots\ .\label{FF}\end{align}\item The vector and
pseudoscalar \emph{decay constants\/} $f_{V,P}$ govern the
meson--vacuum matrix elements of~$j_\mu^{(5)}$:\begin{equation}
\langle0|j_\mu|V(q)\rangle=f_V\,M_V\,\varepsilon_\mu(q)\ ,\qquad
\langle0|j_\mu^5|P(q)\rangle={\rm i}\, f_P\,q_\mu\ .\label{Eq:f}
\end{equation}\end{itemize}In terms of the above quantities, the
contributions of the \emph{poles\/} residing at the masses
$M_{V_R}$ or $M_{P_R}$ of~the relevant vector and pseudoscalar
resonances, $V_R$ and $P_R,$ to the form factors introduced in
Eqs.~(\ref{FF}) read\begin{align*}F_+^{P\succ P'}(q^2)&=
\frac{g_{PP'V_R}\,f_{V_R}}{2\,M_{V_R}}\,\frac{1}{1-q^2/M_{V_R}^2}+
\cdots\ ,\qquad\!V^{P\succ
V}(q^2)=\frac{(M_V+M_P)\,g_{PVV_R}\,f_{V_R}}{2\,M_{V_R}}\,
\frac{1}{1-q^2/M^2_{V_R}}+\cdots\ ,\\A_0^{P\succ V}(q^2)&=
\frac{g_{PP_RV}\,f_{P_R}}{2\,M_{V}}\,\frac{1}{1-q^2/M^2_{P_R}}
+\cdots\ .\end{align*}

\section{Dispersion formalism relying on relativistic
constituent-quark framework}\label{Sec:D}For the actual
theoretical computation of the three transition form factors
${\cal F}(q^2)$ lying at the core~of our strong-couplings study,
past experience leads us to trust in the relativistic
\emph{constituent-quark\/} model~\cite{CQM}. Adhering to this
conviction requires us to match our currents to those built up by
constituent~quarks,~$Q.$\begin{itemize}\item For heavy-quark
currents, this can be easily accomplished by use of corresponding
form factors~$g_{V,A}$:$$j_\mu=g_V\,\bar Q_1\,\gamma_\mu\,Q_2
+\cdots\ ,\qquad j_\mu^5=g_A\,\bar Q_1\,\gamma_\mu\,\gamma_5\,Q_2
+\cdots\ .$$\item For light-quark currents, partial axial-current
conservation, e.g., renders this rather cumbersome~\cite{GVA}.
\end{itemize}Following Ref.~\cite{MS}, we use for our model
parameter values the constituent-quark masses of
Table~\ref{Tab:m}~and$$g_V=g_A=1\ .$$

\begin{table}[hb]\centering\caption{Numerical values of the
constituent-quark mass parameters $m_Q$ entering in the present
investigation \cite{MS}.}\label{Tab:m}\begin{tabular}{lcccc}
\toprule Quark flavour $Q$&$u$&$d$&$s$&$c$\\\midrule Quark mass
$m_Q$\;\mbox{(GeV)}&0.23&0.23&0.35&1.45\\\bottomrule\end{tabular}
\end{table}

\begin{figure}[hbt]\centering
\includegraphics[scale=.2,clip]{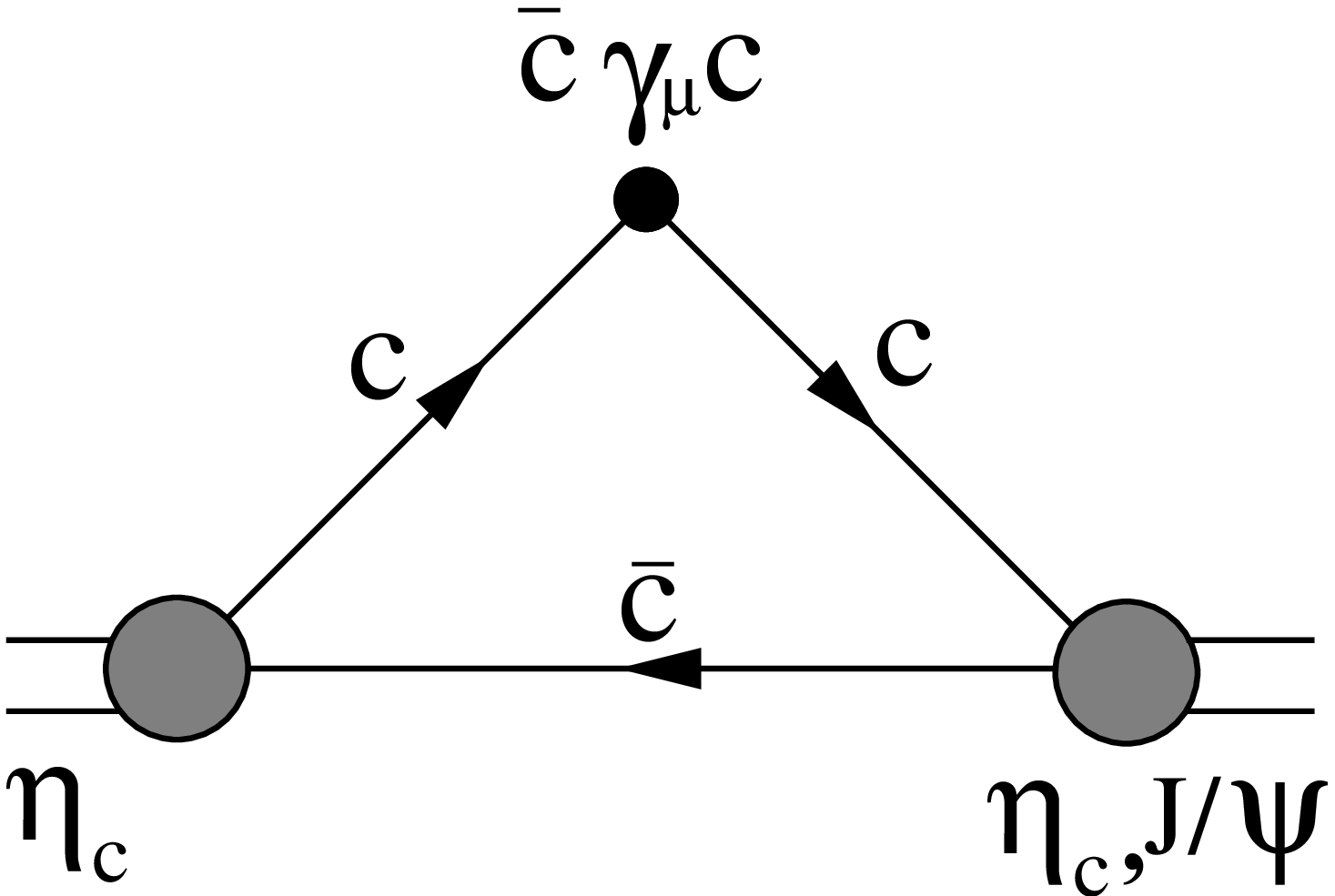}\hspace{10ex}
\includegraphics[scale=.2,clip]{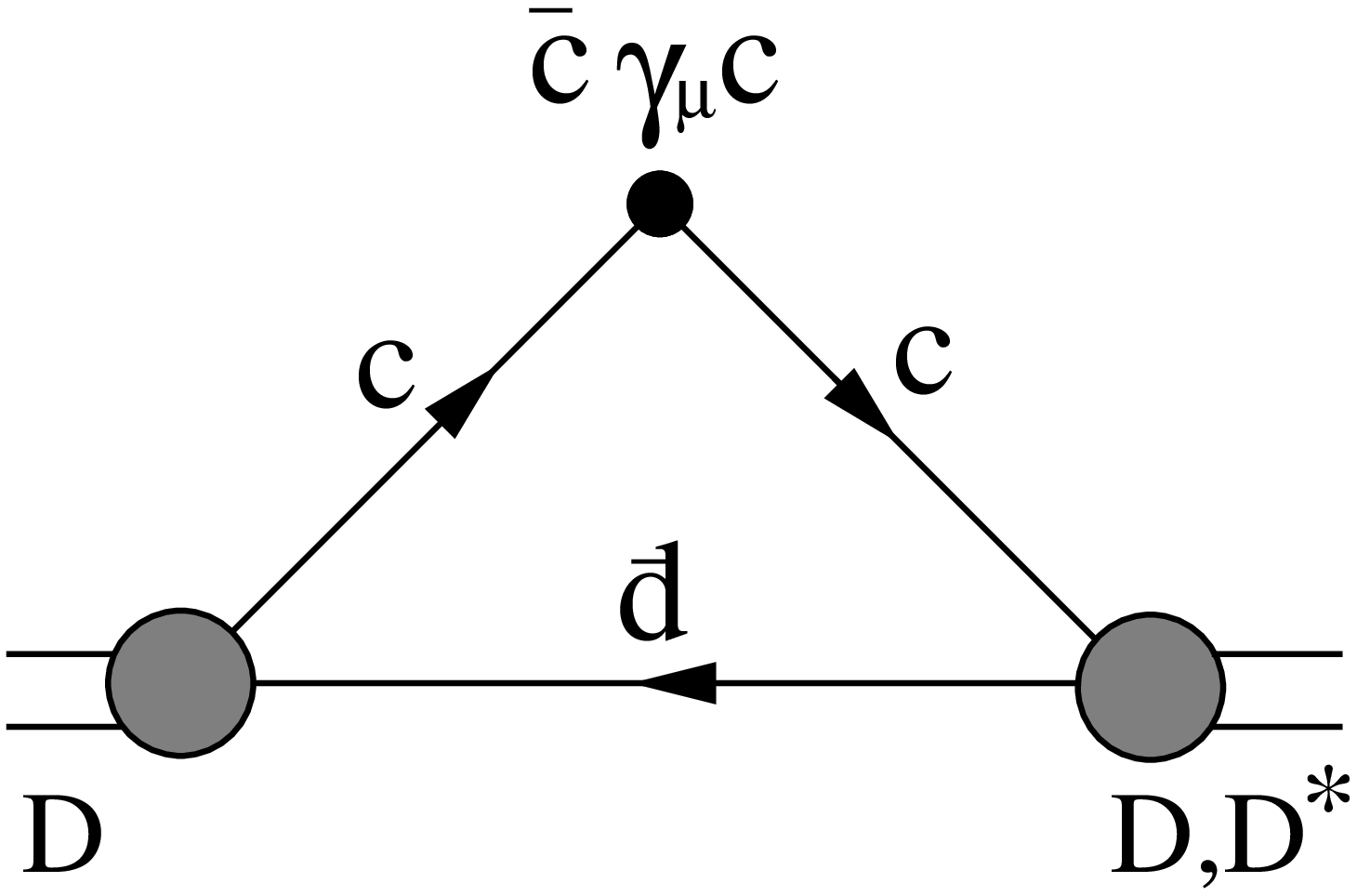}\hspace{10ex}
\includegraphics[scale=.2,clip]{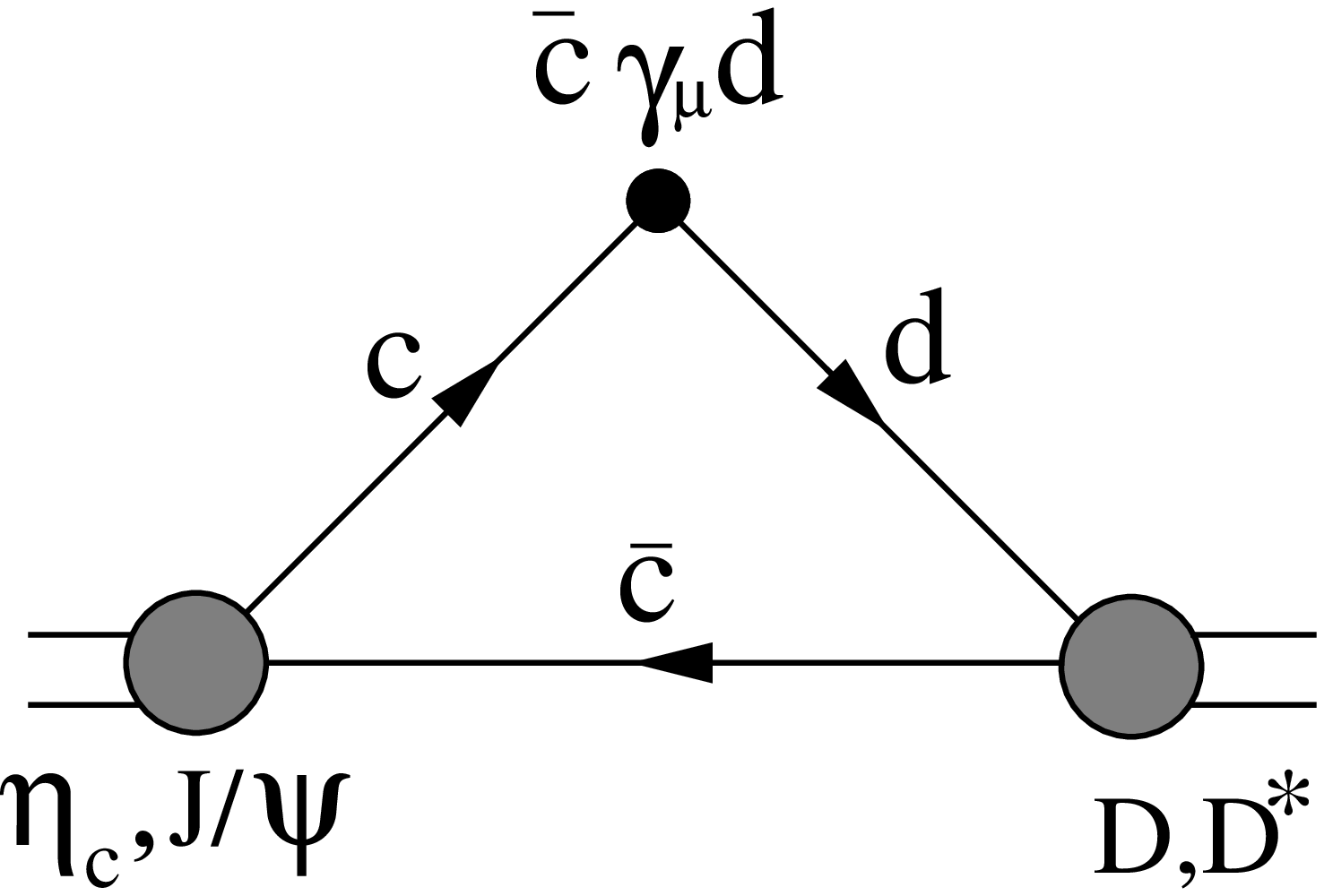}\caption{$M_1\succ M_2$
meson--meson transitions for $M_{1,2}=\eta_c,J/\psi,D,D^*,$
mediated by the \emph{constituent\/}-quark vector currents $\bar
c\,\gamma_\mu\,c$ or $\bar c\,\gamma_\mu\,d$: Feynman graphs
yielding the one-loop contributions to the spectral density
$\Delta_{\cal F}(s_1,s_2,q^2)$.}\label{Fig:FD}\end{figure}

Then, by application of the relativistic dispersion formalism, we
are put in the position to represent the leptonic decay constants
$f_{P,V}$ in the form of dispersion integrals of \emph{spectral
densities\/} $\rho_{P,V}(s)$ and the transition form factors
${\cal F}(q^2)$ by double dispersion integrals of \emph{double
spectral densities\/} $\Delta_{\cal F}(s_1,s_2,q^2),$
\begin{equation}f_{P,V}=\int{\rm d}s\,\phi_{P,V}(s)\,\rho_{P,V}(s)\
,\qquad{\cal F}(q^2)=\!\int{\rm d}s_1\,\phi_1(s_1)\!\int{\rm
d}s_2\,\phi_2(s_2)\,\Delta_{\cal F}(s_1,s_2,q^2)\
,\label{Eq:fF}\end{equation}involving the \emph{wave functions\/}
of pseudoscalar and vector mesons taking part in the studied
reactions~\cite{RDA}$$\phi_{P,V}(s)=\frac{\pi}{s^{3/4}}
\sqrt{\frac{s^2-(m_1^2-m^2)^2}{2\,[s-(m_1-m)^2]}}\,w_{P,V}\!
\left(\frac{(s-m_1^2-m^2)^2-4\,m_1^2\,m^2}{4\,s}\right)\ ,\qquad
\int{\rm d}k\, k^2\,w^2_{P,V}(k^2)=1\ .$$The spectral densities
may be derived from one-loop Feynman graphs, of the kind shown in
Fig.~\ref{Fig:FD}. For the radial meson wave functions
$w_{P,V}(k^2),$ it has become customary to assume simple
Gaussian~shapes:\begin{equation}w_{P,V}(k^2)\propto\exp
\left(-\frac{k^2}{2\,\beta_{P,V}^2}\right)\ .\label{Eq:w}
\end{equation}The necessary input parameter values, drawn from a
variety of related sources, are collected in Table~\ref{Tab:M}.
\pagebreak

\begin{table}[hbt]\centering\caption{Numerical values of the
relevant parameters of the charmed mesons $D_{(s)}^{(*)}$ and the
charmonia $\eta_c$ and $J/\psi$: meson mass $M,$ leptonic decay
constant $f$ defined in Eq.~(\ref{Eq:f}), and slope $\beta$ fixing
the width of the Gaussian (\ref{Eq:w}) \cite{MP}.}\label{Tab:M}
\begin{tabular}{ccccccc}\toprule
Meson&$D$&$D^*$&$D_s$&$D_s^*$&$\eta_c$&$J/\psi$\\\midrule
$M$~(GeV)\ &$1.87$&$2.010$&$1.97$&$2.11$&$2.980$&$3.097$\\[.5ex]
$f$~(MeV)\ &\ $206\pm8$\ &\ $260\pm10$\ &\ $248\pm2.5$\ &\
$311\pm9$\ &\ $394.7\pm2.4$\ &\ $405\pm7$\ \\[.5ex]$\beta$~(GeV)\
&$0.475$&$0.48$&$0.545$&$0.54$&$0.77$&$0.68$\\\bottomrule
\end{tabular}\end{table}

In order to extract the strong couplings under discussion, we
determine the momentum dependence for the transition form factors
${\cal F}(q^2)$ sufficiently far off their resonances $R,$ where
$R=V$ for ${\cal F}=F_+^{P\succ P'}$ and ${\cal F}=V^{P\succ V}$
but $R=P$ for ${\cal F}=A_0^{P\succ V},$ \emph{interpolate\/} our
results by means of the simple parametrization\begin{equation}
{\cal F}(q^2)=\frac{{\cal F}(0)}{\left(1-q^2/M_R^2\right)
\left(1-\sigma_1\,q^2/M_R^2+\sigma_2\,q^4/M_R^4\right)}\ ,\qquad
\mbox{Res}\,{\cal F}(M_R^2)=\frac{{\cal F}(0)}
{1-\sigma_1+\sigma_2}\ ,\label{Eq:F}\end{equation}governed by the
three parameters $\sigma_{1,2}$ and ${\cal F}(0),$ and extrapolate
this momentum dependence of ${\cal F}(q^2)$ to the resonance
region $q^2\approx M_R^2.$ From the resulting \emph{residues\/} of
the meson poles at $q^2=M_R^2,$ the~strong couplings are found by
factorizing off all known quantities such as meson masses and
decay constants. In case a particular strong coupling shows up in
residues of resonance poles of more than one transition form
factor, for such multipresent strong coupling an optimized
estimate is obtained by a combined~fit.

\section{\boldmath Strong couplings among three $\eta_c$ or
$J/\psi$ mesons: $\eta_c\eta_cJ/\psi$ and $\eta_cJ/\psi J/\psi$}
\label{Sec:c}An illustration of such \emph{multipresence\/} is
given by the strong coupling $g_{\eta_c\eta_c\psi}$ \cite{LMSS},
with appearance in~both\begin{itemize}\item the residue of
$F_+^{\eta_c\succ\eta_c}(M_\psi^2)$ arising from the vector
current $\bar c\,\gamma_\mu\,c$ coupling, with strength $f_\psi,$
to $J/\psi$~and\item the residue of
$A_0^{\eta_c\succ\psi}(M_{\eta_c}^2)$ from the axial-vector
current $\bar c\,\gamma_\mu\,\gamma_5\,c$ that couples, with
strength $f_{\eta_c},$ to~$\eta_c$:\end{itemize}$$\mbox{Res}\,
F_+^{\eta_c\succ\eta_c}(M_\psi^2)=g_{\eta_c\eta_c\psi}\,
\frac{f_\psi}{2\,M_\psi}\ ,\qquad\mbox{Res}\,A_0^{\eta_c\succ\psi}
(M_{\eta_c}^2)=g_{\eta_c\eta_c\psi}\,\frac{f_{\eta_c}}{2\,M_\psi}\
.$$

\begin{figure}[hbt]\centering
\includegraphics[scale=.4006,clip]{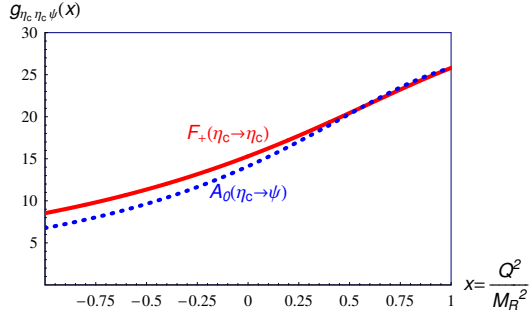}
\caption{Off-shell strong coupling $g_{\eta_c\eta_c\psi}(x)$ as
function of $x\equiv\frac{q^2}{M_R^2}$ for $\eta_c\succ\eta_c$
(\textcolor{red}{red}) or $\eta_c\succ J/\psi$
(\textcolor{blue}{blue}) transitions.}\label{Fig:c}\end{figure}

After such detailed preliminaries, the way how to proceed should
be pretty plain: We determine the meson wave-function parameters
$\beta_{\eta_c,\psi}$ of both charmonia such that the dispersion
representation~(\ref{Eq:fF})~of their decay constants,
$f_{\eta_c,\psi},$ reproduces the observed values. With these
meson vertices at our disposal, we deduce the strong coupling of
interest, for each meson--meson transition where this strong
coupling enters (for the case of the $\eta_c\eta_cJ/\psi$
coupling, see Fig.~\ref{Fig:c}), from the spectral representation
(\ref{Eq:fF}) of the form factor corresponding to the respective
transition. Thus, our $\eta_c$ and $J/\psi$ $PPV$ and $PVV$
couplings~are~\cite{LMSS}$$g_{\eta_c\eta_c\psi}=25.8\pm1.7\
,\qquad g_{\eta_c\psi\psi}=(10.6\pm1.5)\;{\rm GeV}^{-1}\ .$$

\section{\boldmath Strong three-meson couplings of the charmonia
$J/\psi$ or $\eta_c$ to $D_{(s)}$ and $D^*_{(s)}$}\label{Sec:d}
\begin{figure}[hbt]\centering\begin{tabular}{ccc}
\includegraphics[scale=.466,clip]{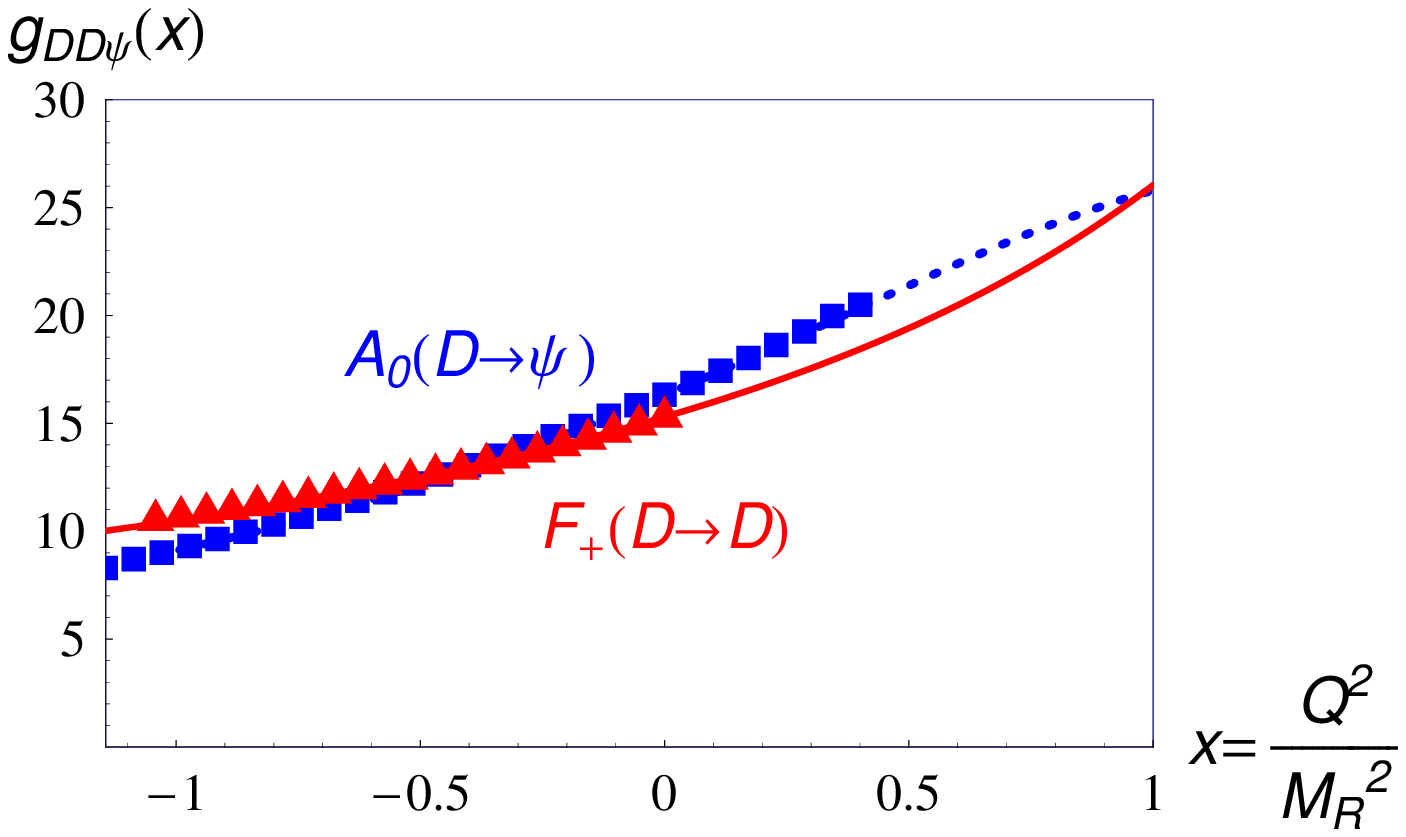}&
\includegraphics[scale=.409,clip]{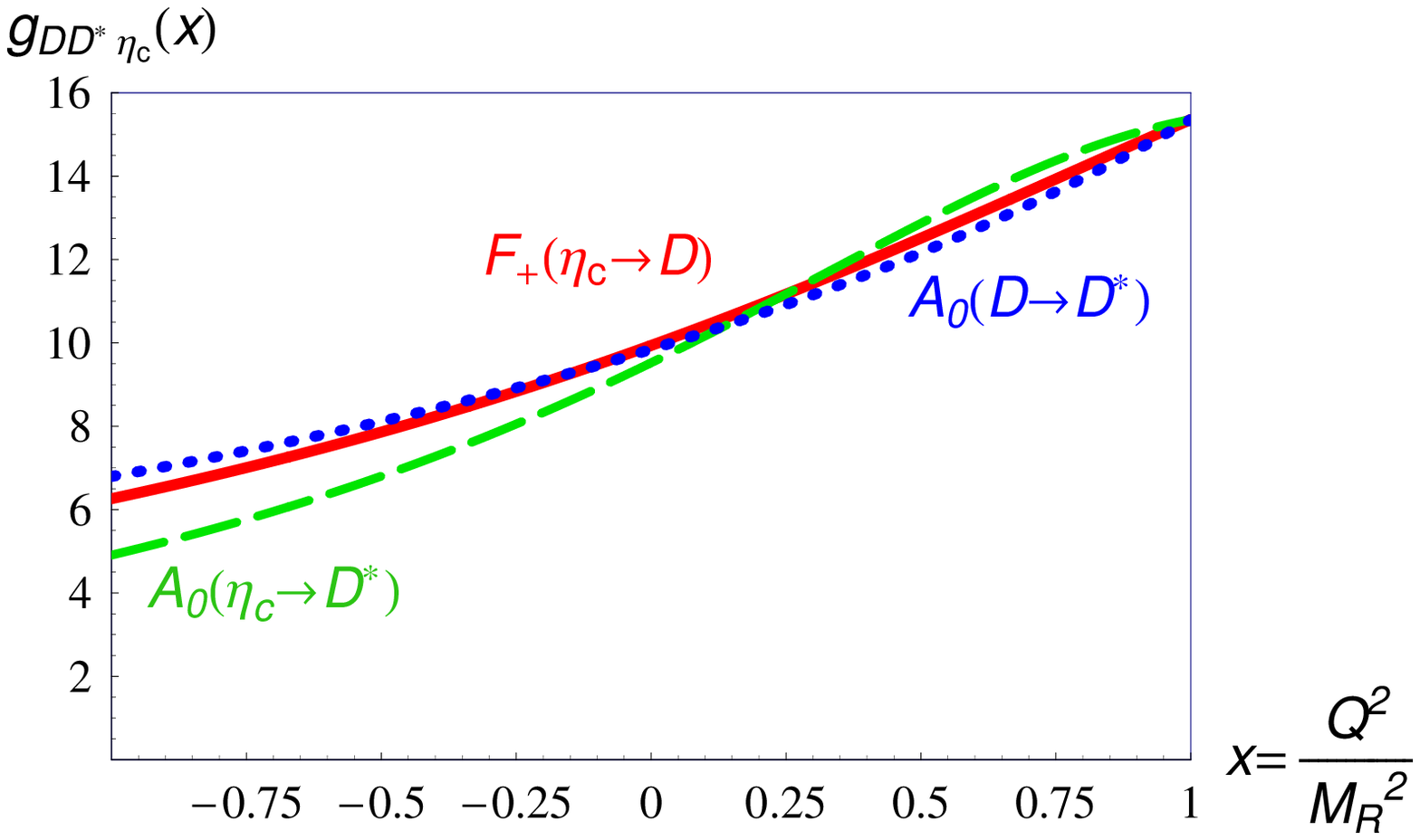}\\(a)&(b)
\end{tabular}\caption{Off-shell strong couplings $g_{DD\psi}$ and
$g_{DD^*\eta_c}$ to $D^{(*)}$ mesons (with resonances indicated by
circumflexes): dependences on $x\equiv\frac{q^2}{M_R^2}$ of (a)
$g_{D\hat D\psi}(x)=\frac{2\,M_\psi}{f_D}\,(1-x)\,
A^{D\succ\psi}_0(q^2)$ (\textcolor{blue}{blue}) and
$g_{DD\hat\psi}(x)=\frac{2\,M_\psi}{f_\psi}\,(1-x)\,F_+^{D\succ
D}(q^2)$ (\textcolor{red}{red}),~got with (lines) and without
(symbols) interpolation, and of (b) $g_{D\hat{D^*}\eta_c}(x)$
(\textcolor{red}{red}), $g_{DD^*\hat{\eta_c}}(x)$
(\textcolor{blue}{blue}) and $g_{\hat DD^*\eta_c}(x)$
(\textcolor{green}{green}).}\label{Fig:d}\end{figure}

Along the same route as in Sec.~\ref{Sec:c} --- and by taking into
account, in addition, constituent-quark currents involving one $d$
or $s$ quark --- we may likewise discuss the strong couplings of
$J/\psi$ or $\eta_c$ to non-strange ($D^{(*)}$) or strange
($D_s^{(*)}$) charmed mesons. Combining the options sketched in
Fig.~\ref{Fig:d}, our findings are~\cite{LMSS}\begin{alignat}
{2}g_{DD\psi}&=26.04\pm1.43\ ,\qquad&g_{DD^*\psi}&=
(10.7\pm0.4)\;{\rm GeV}^{-1}\ ,\nonumber\\g_{DD^*\eta_c}&=
15.51\pm0.45\ ,\qquad&g_{D^*D^*\eta_c}&=(9.76\pm0.32)\;{\rm
GeV}^{-1}\ ,\nonumber\\g_{D_sD_s\psi}&=23.83\pm0.78\
,\qquad&g_{D_sD_s^*\psi}&= (9.6\pm0.8)\;{\rm GeV}^{-1}\
,\nonumber\\g_{D_sD_s^*\eta_c}&=14.15\pm0.52\
,\qquad&g_{D_s^*D_s^*\eta_c}&=(8.27\pm0.37)\;{\rm GeV}^{-1}\
.\label{Eq:d}\end{alignat}

\section{Observations, comparison with findings of different
origin, conclusions}\label{Sec:S}Our application of a relativistic
dispersion technique, relying on the constituent-quark model, to
strong three-meson couplings of quarkonia among each other and to
$D^{(*)}_{(s)}$ mesons yields some crucial insights:
\begin{enumerate}\item The interpolation of our numerical
transition-form-factor results found at low $q^2$ by means of the
ansatz (\ref{Eq:F}) yields values of the resonance-mass fit
parameter $M_R$ close to the experimental~meson masses; this can
be interpreted as confirmation of the presence of the poles
expected at $q^2\approx M_R^2.$\item The replacement of the $d$
quark by the $s$ quark (or vice versa) in the quark currents
mediating any transition under study enables us to arrive at some
estimate of the size of SU(3)-breaking~effects. Inspecting
Eq.~(\ref{Eq:d}), we get a change of the strong couplings under
consideration by roughly 10\%.\item Table~\ref{Tab:C} confronts,
for the strong couplings between \emph{charmonia and
$D_{(s)}^{(*)}$ mesons\/}, the predictions of our dispersive
constituent-quark formalism with (available) corresponding figures
from QCD sum rules \cite{Mat05,Bra14,Bra15}; surprisingly, the
latter prove to be smaller than our results \cite{LMSS} by a
factor of~two.\end{enumerate}

\begin{table}[hb]\centering\caption{Strong couplings of three
mesons which involve one $J/\psi$: findings of the present
relativistic constituent quark-model framework \cite{LMSS},
confronted with available results \cite{Mat05,Bra14,Bra15}
extracted from the QCD sum-rule approach.}\label{Tab:C}
\begin{tabular}{lcccc}\toprule Coupling&$g_{DD\psi}$&$g_{DD^*\psi}$
(GeV$^{-1}$)&$g_{D_sD_s\psi}$&$g_{D_sD_s^*\psi}$ (GeV$^{-1}$)\\
\midrule Quark model \cite{LMSS}&$26.04\pm1.43$& $10.7\pm0.4$&
$23.83\pm0.78$&$9.6\pm0.8$\\[.5ex]QCD sum rules&$11.6\pm1.8$
\cite{Mat05}&$4.0\pm0.6$ \cite{Mat05}&$11.96\pm1.34$ \cite{Bra14}&
$4.30\pm1.53$ \cite{Bra15}\\\bottomrule\end{tabular}\end{table}

\end{document}